\documentclass[letterpaper,prl,twocolumn,floatfix,superscriptaddress]{revtex4}  
\usepackage[normalem]{ulem}  
\usepackage{amsmath}
\usepackage[mathcal]{euscript}
\usepackage{mathrsfs}
\usepackage{float}
\usepackage[usenames,pdftex]{color}
\usepackage[pdftex]{color,graphicx}     
\usepackage[pdftex,bookmarks=false]{hyperref}           

\begin{document}

\title{On the surface of glasses}
\date{\today}
\author{Jacob D. Stevenson}
\affiliation{Department of Physics and Department of Chemistry and Biochemistry,
Center for Theoretical Biological Physics,
University of California, San Diego, La Jolla, CA 92093}
\author{Peter G. Wolynes}
\affiliation{Department of Physics and Department of Chemistry and Biochemistry,
Center for Theoretical Biological Physics,
University of California, San Diego, La Jolla, CA 92093}
\affiliation{e-mail: pwolynes@ucsd.edu}

\begin{abstract} Dynamics near the surface of glasses is generally much faster
  than in the bulk.  Neglecting static perturbations of structure at the
  surface, we use random first order transition theory to show the free energy
  barrier for activated motion near a free surface should be half that of the
  bulk at the same temperature.  The increased mobility allows the surface
  layers to descend much further on the energy landscape than the bulk
  ordinarily does.  The simplified RFOT calculation however predicts a limiting
  value for the configurational entropy a vapor deposited glass may reach as a
  function of deposition rate.  We sketch how mode coupling effects extend the
  excess free surface mobility into the bulk so that the glass transition
  temperature is measurably perturbed at depths greater than the naive length
  scale of dynamic cooperativity.  \end{abstract}

\maketitle
\bibliographystyle{/people/jake/latex_files/naturemag}

Paying attention to the properties of interfaces has, since the time of Van der
Waals, helped to clarify our understanding of bulk phase
transitions\cite{rowlinson.1982}.  Studying the mobility at the surface of
glasses and supercooled liquids has the potential to be equally enlightening
about the glass transition\cite{swallen.2007,fakhraai.2008}.  Theories of the
glass transition predict a growing length scale of correlated dynamics as the
glass transition is
approached\cite{lubchenko.2007,kivelson.1995,franz.2000,biroli.2004}.  This
growth, according to the most successful theories, is quite modest.  It has
only in recent years been widely acknowledged that there is such a slowly
growing length scale in bulk glasses\cite{berthier.2005,capaccioli.2008}.  The
dynamics of a glass or supercooled liquid should be perturbed within a few
correlation lengths of its surface.  Many experiments do show significant
perturbations of the glassy dynamics at free surfaces or in confined
spaces\cite{alcoutlabi.2005}.  The picture that emerges from these experiments
is, at present, still somewhat confused.  In general, mobility seems to be
increased at a free surface, although sometimes a diminished mobility has been
observed.  At interfaces with solids both increases and decreases in mobility
have been reported\cite{forrest.1996,li.2005}.  It seems likely that this
complex set of behaviors reflects the fact that the dynamics of the liquid can
be strongly influenced by the liquid's static structure which also will be
perturbed by the interface --- in the extremes, partial crystallization can
occur at a free interface while a drying layer may sometimes insulate a
confined fluid from its solid surroundings.  While noting these complications,
we feel it is nevertheless worthwhile to present a simplified view of glass and
supercooled liquid surfaces in the context of the random first order transition
theory of the glass transition.  

The random first order transition (RFOT) theory is a constructive approach to
structural glass dynamics that has explained quantitatively numerous bulk glass
phenomena\cite{lubchenko.2007}.  It was recognized very early on that growing
length scales should be associated with an ideal glass transition as envisaged
in RFOT theory\cite{kirkpatrick.1989,kirkpatrick.1987b}.  Indeed some of the
early experiments on confined supercooled liquids carried out by
Jonas\cite{zhang.1992} were motivated by a desire to test these expectations.
It turns out present day RFOT theory for molecular liquids can make somewhat
more definite predictions than was done in those early days, at least for an
idealized interface which can be taken to have no change in static structure
from the bulk.  We show for this idealization that the maximum mobility at a
completely free interface is related in a very simple way to the bulk mobility.
In terms of relaxation times, the RFOT result for the surface relaxation time
is simply $\tau_{surf} = \sqrt{\tau_0 \tau_{bulk}}$.  We also show that RFOT
theory suggests that as measured by an effective local glass transition
temperature $T_g^{local}$ the influence of the interface can appear to be
rather far-reaching into the bulk, consistent with some experiments.

At the smallest length scales the dynamic arrest of glasses results from the
cage effect in which a particle's motion is constrained by the presence of its
neighbors.  In mean field theory this leads to a friction crisis at the mode
coupling critical temperature and the emergence of an ensemble of aperiodic
crystal structures having an extensive configurational entropy $s_c$ per
particle.  Below the dynamical transition at $T_c$ any large scale motion that
takes place is necessarily collaborative.  Near a free surface this picture is
modified.  Since surface particles feel a weaker structural cage, essentially
only on the inner side, they would go through a dynamic arrest at a lower
temperature and remain more mobile below the bulk glass transition temperature.
The collaborative dynamics propagates the enhanced mobility at the surface some
distance into the bulk.  This depth would be determined by the length scale of
cooperativity.

In the mean field limit (made precise in Kac models by Franz\cite{franz.2007})
two cooperative length scales can be defined for random first order transition
theory\cite{kirkpatrick.1987b}.  One scale, $\xi_{MCT}$, is directly related to the dynamical
transition at $T_c$ which resembles a spinodal\cite{biroli.2004}, the other is
related to the size of regions that rearrange by activated motions and scales
with the configurational entropy $s_c$\cite{xia.2000}.  The latter scale
diverges at the Kauzmann temperature with an exponent $\xi_{RFOT} \sim a
s_c^{-2/3}$.  Within the sharp, thin wall approximation the RFOT theory
yields a numerical coefficient $a$ that appears consistent with experimental
values for molecular glasses\cite{xia.2000}.  The scaling exponent used to fit
inferred lengths from $\chi_4$ is also not inconsistent with RFOT theory,
although a larger value provides a better
fit\cite{berthier.2005,capaccioli.2008}.  For molecular liquids governed by
short range interactions the dynamical (mode coupling) length would not
actually diverge at $T_c$ because of a dynamical cutoff from the activated
events and will not be too different from $\xi_{RFOT}$.

We first review the theory of bulk activated dynamics\cite{lubchenko.2007}.
Below the mode coupling theory (MCT) transition, dynamics takes place on a
rugged free energy landscape\cite{kirkpatrick.1987b,mezard.1999,singh.1985}.
Particle motion occurs locally through transitional hops between metastable
structural states which resemble, in many ways, nucleation processes between
different aperiodic crystal structures\cite{xia.2000,kirkpatrick.1987b}.  The
magnitude of the free energy barrier can be found by a competition between the
entropic cost of remaining confined to one minimum free energy structure and a
mismatch free energy penalty, $\sigma$, for having two mean field solutions
adjacent to each other but in distinct structural states.  The free energy
profile governing nucleation of a spherical cooperative region of radius $r$ is

\begin{equation}
  F(r) = 4 \pi r^2 (r_0 / r)^{1/2} \sigma_0 - \frac{4}{3} \pi r^3 n_0 T s_c
  \label{eqn:nucleation1}
\end{equation}

The surface penalty $\sigma(r)= (r_0 / r)^{1/2} \sigma_0$ scales with $r$ due
to renormalization effects of wetting by the diverse set of
structures\cite{xia.2000,villain.1985}.  The $r^{-1/2}$ scaling holds
sufficiently close to the ideal glass transition.  At short length scales
$\sigma_0$ can be obtained from a crude density functional calculation giving
$\sigma_0 = \frac{3}{4} k_B T r_0^{-2} \ln \frac{1}{d_L^2 \pi e}$ where $d_L
\approx 0.1$ particle spacings is the Lindemann length,  $n_0 = r_0^{-3}$ is
the density of the liquid. The configurational entropy, $s_c$, measures the
number of available structural states.  The resulting free energy barrier for
reconfiguration events in bulk is

\begin{equation}
  F^{\ddagger} = \frac{3 \pi \sigma_0^2 r_0^4}{T s_c(T)}
  \label{eqn:fdagbulk}
\end{equation}

The relaxation time $\tau_{\alpha} = \tau_0 e^{F^{\ddagger}/k_BT}$ diverges at
the Kauzmann temperature ($s_c(T_K) \to 0$) and thus follows the Vogel-Fulcher
behavior  in the deeply supercooled region $\tau_{\alpha} \sim e^{B/(T-T_K)}$.
The length scale of an activated event also follows from the free energy
profile

\begin{equation}
  r^{*} = r_0 \left( \frac{3 \sigma_0 r_0^{2} }{T s_c(T)} \right)^{2/3}
  \label{eqn:rstar}
\end{equation}

\noindent This length scale increases with decreasing temperature, reaching a
universal value of about 5 inter-particle spacings at the glass
transition temperature corresponding to a one hour relaxation time.

\begin{figure}[tp]
  \begin{center}
    \includegraphics[width=0.48\textwidth]{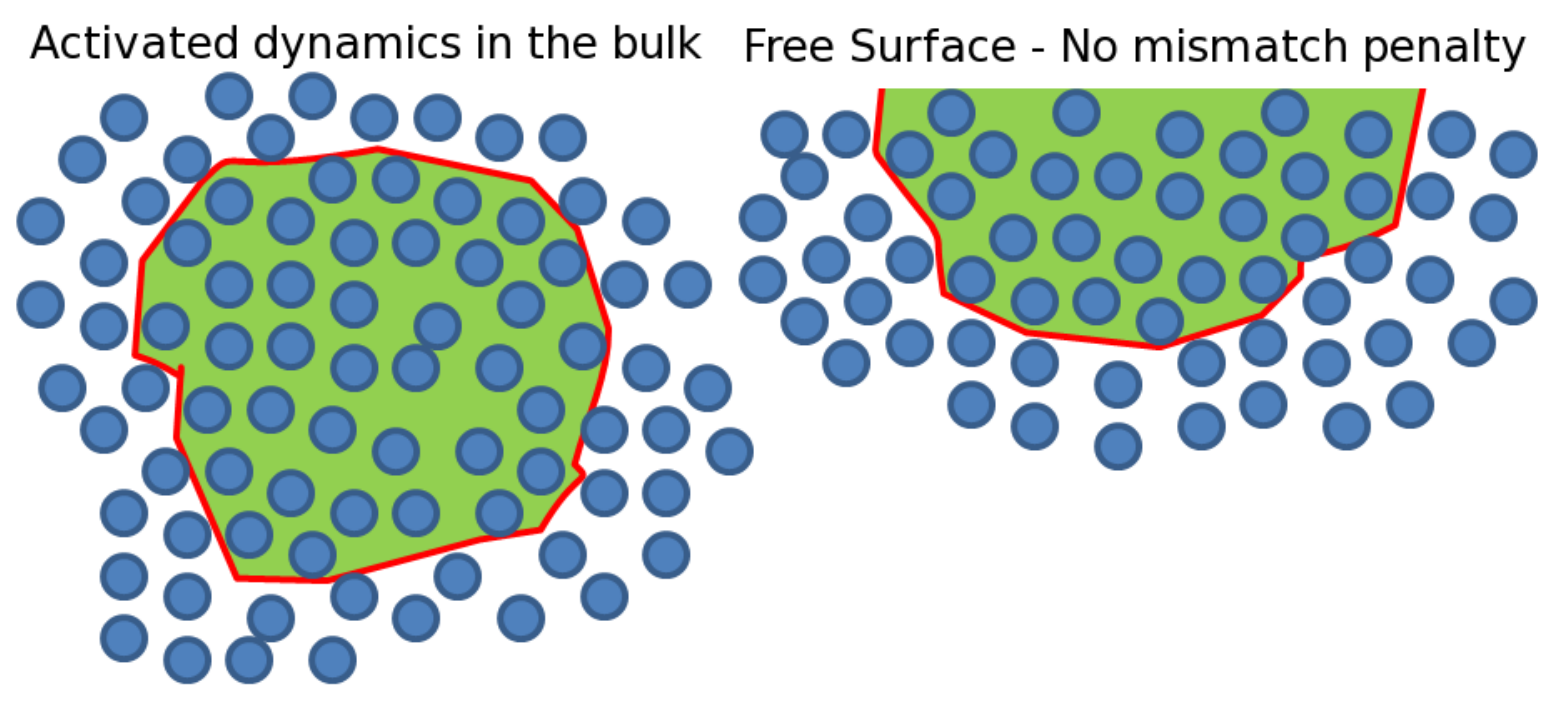}
  \end{center}
  \caption{In the bulk, at low temperatures, activated motion occurs within
  roughly spherical regions.  Near a free surface a rearranging hemispherical
  region feels no mismatch penalty on interface along the free surface leading
  to much faster dynamics than the bulk.  }
  \label{fig:hemisphere}
\end{figure}

The RFOT analysis of bulk activated dynamics can be easily modified (see figure
\ref{fig:hemisphere}) to treat motions near a free surface.  Near a completely
free surface with no structural modifications, the transition state would
rearrange a region of hemispherical shape in order to minimize the surface area
subject to the mismatch penalty.  This shape can be shown to be appropriate
when wetting effects are neglected\footnote{For shapes other than hemispheres
the wetting effect would need to be modified to correctly deal with the
curvature, but the hemisphere, or something close to it, would still be the
most favored shape.}.  The flat face of the hemisphere lies along the free
surface and is assumed not to contribute at all to the mismatch penalty.  The
free energy profile for the hemispherical rearranging region (including
wetting) becomes

\begin{equation}
  F_{surf}(r) = 2 \pi r^2 (r_0 / r)^{1/2} \sigma_0 - \frac{2}{3} \pi r^3 n_0 T s_c.
  \label{eqn:nucleation2}
\end{equation}

\noindent Since both terms in the expression are halved from the equation for
bulk dynamics we see that the size corresponding to the transition state for
activated rearrangement remains unchanged from that for the bulk, but the
resulting free energy barrier is reduced by a factor of 2,

\begin{equation}
  F^{\ddagger}_{surf} = \frac{1}{2} \frac{3 \pi \sigma_0^2 r_0^4}{T s_c(T)}
  \label{eqn:fdagsurf}
\end{equation}

\noindent This seemingly innocuous change leads to dramatically faster
relaxation at free surfaces than occurs in the bulk.  Molecular motion near the
surface of a glass would still be detectable even if the bulk were essentially
frozen.  We see that very near the surface, the relaxation time is related very
simply to the bulk value,

\begin{equation}
  \tau_{surf} = \sqrt{\tau_0 \tau_{bulk}}.
\end{equation}

According to RFOT, on laboratory time scales, the ideal surface layer will be
able to ``hike'' (using the colorful expression of Ediger\cite{kearns.2008})
much further down the free energy landscape than expected, reaching lower
energy, more stable configurations than the bulk is likely to find.  The
surface motion will freeze out, i.e. go through its own glass transition
($T_g^{surf}$) only when $\tau_{surf}$ becomes larger than the laboratory time
scale.  For the same laboratory time scale, according to equation
\ref{eqn:fdagsurf} this will happen when the configurational entropy is half
the characteristic configurational entropy of the bulk glass transition. The
bulk glass transition entropy is very nearly material
independent\cite{lubchenko.2003,stevenson.2005}, and likewise the
free surface glass transition will occur at a characteristic \emph{bulk}
entropy value.  For a one hour time scale this characteristic entropy is

\begin{equation}
  s_c(T_g^{surf}) = \frac{1}{2} s_c(T_g) \approx 0.41k_B
\end{equation}

Ediger and co-workers\cite{kearns.2008} have recently shown how to construct a
microscopic glass sample via vapor deposition reaching much lower in the energy
landscape than was possible by the usual bulk cooling of a liquid.  They argue this
ultra-stable glass arises from the free surface's excess mobility.  For slow
deposition on a cold substrate the growing glass, owing to the higher mobility,
can more thoroughly rearrange near the free surface to form exceedingly stable
structures.  They assign a fictive temperature to their sample, $T_f$, defined
as the temperature where a glass of similar stability, but created by bulk
cooling would fall out of equilibrium.  $T_f$ can be used to measure the
sample's position on the free energy landscape relative to an ordinary glass
through the parameter,

\begin{equation}
  \theta_K = \frac{T_g - T_f}{T_g - T_K}
  \label{eqn:theta}
\end{equation}

\noindent This parameter increases with the stability of the glass, reaching one
for an ideal glass.  In principle if there were in fact no entropy crisis,
$\theta_K$ could exceed one.  $\theta_K$ can be parametrized by the configuration
entropy, an equivalent measure of structural stability.  Using the linear
extrapolation vanishing at $T_K$ valid for low
temperature ($T<T_g$) the RFOT free surface mobility would yield, for a rate of
deposition equal to one correlation length per hour

\begin{equation} \theta_K = 1 - \frac{s_c(T_f)}{s_c(T_g)} \le \frac{1}{2}
\end{equation}

\noindent As the enhanced surface mobility freezes out at $T_g^{surf}$, this
temperature should be seen as a limiting value for $T_f$ leading to the
inequality above.  Ediger's experiments yielded $\theta_K \approx 0.4$ for both
liquids studied\cite{kearns.2008}.  It would take $10^4$ years to to make these
highly stable glasses via traditional bulk cooling, while $10^9$ years, roughly
the age of the universe, would be required for a glass with the limiting
stability of $\theta_K=1/2$.  Unless the surface structure is greatly modified
leading to a locally smaller $T_K$, vapor deposition should not yield glasses
with stability much greater than the naive RFOT limiting value.

The linear extrapolation for $s_c(T)$ is valid near $T_K$, but for some
liquids, at higher temperatures a better approximation is\cite{richert.1998}
$s_c = s_{\infty} (1-T_K/T)$.  Recognizing this feature does add some material
dependence to the limiting stability

\begin{equation}
  \theta_K \le \frac{T_g}{T_g + T_K}.
\end{equation}

\noindent Fragile liquids still have a limit near $1/2$, but strong liquids
have the possibility of descending further in terms of temperature (but
\emph{not} in terms of configurational entropy) than fragile liquids.  The fragilities for
the two glasses so far studied by Ediger are similar and would yield $\theta_K
\le 0.57$(IMC) and $\theta_K \le 0.58$(TNB) for the nanometer per hour
deposition rate.  

A key prediction of this analysis is that the fictive temperature is a
logarithmic function of the deposition rate.  More precisely the
configurational entropy at the fictive temperature is a universal function of
the logarithmic deposition rate.  The limiting fictive temperature given above
assumes a deposition rate of one correlation length per hour.  More generally,
for a deposition rate $k$ the mobile surface layer of depth $\xi$ will be
equilibrated on a time scale $\xi / k$, which is related to the fictive
temperature through the relationship $\xi / k = \tau_0 e^{F^{\ddagger}(T_f) /
k_B T_f} =   \tau_0 e^{A' / s_c(T_f)}$.  

\begin{equation}
  s_c(T_f) = \frac{A' \log e}{\log \xi / k \tau_0}
\end{equation}

\begin{figure}[tp]
  \begin{center}
    \includegraphics[width=0.48\textwidth]{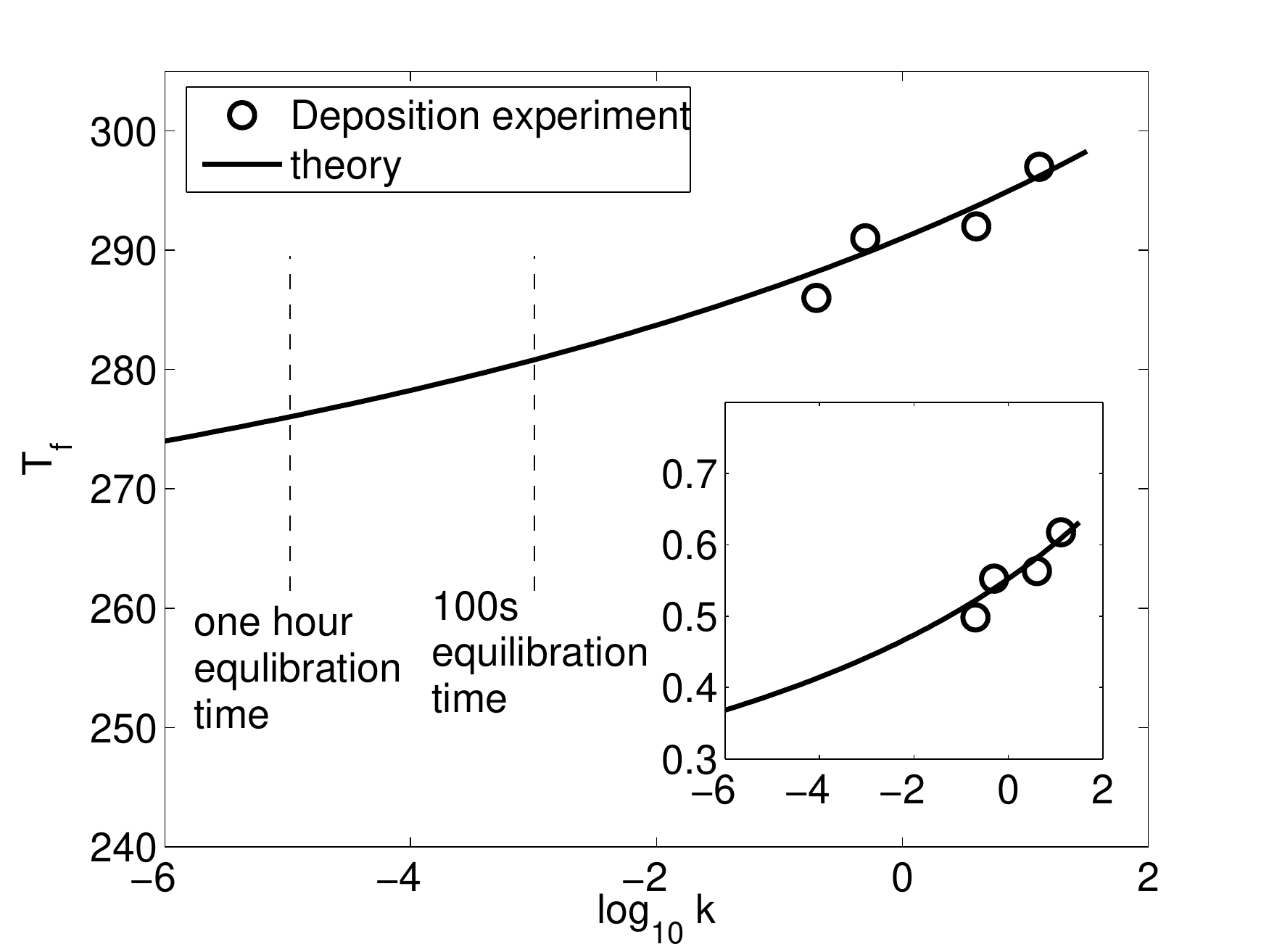}
  \end{center}
  \caption{Fictive temperatures vs deposition rate for the glass former IMC.
  Data for the deposition experiment was taken from reference
  \protect{\cite{kearns.2008}}.  The same data are shown in the inset plotted as
  $s_c(T_f)$ in Boltzmann units per bead vs deposition rate.}
  \label{fig:fictivetemp}
\end{figure}

We compared to Ediger's experimental results using the linear representation
for the configurational entropy, $s_c(T_f) = \Delta C_p(T_g) (T_f - T_K) /
T_K$.  Along with $A'$ obtained from equation \ref{eqn:fdagsurf}, we take $\xi
\approx 1 nm$, $\tau_0 \sim 10^{-12} s$.  $\Delta C_P(T_G) \approx 2.6 k_B$ is
the heat capacity jump at the glass transition for IMC in Boltzmann units per
bead\cite{stevenson.2005}.  The theoretical results along with the measured
values obtained by Ediger et al.\cite{kearns.2008} are plotted in figure
\ref{fig:fictivetemp}.  Our results are consistent with the experimental data
for the deposition rates tested.  The theory correctly predicts the trend for
slower deposition rates.  Despite the good agreement this is a zeroth order
approximation in which we have assumed equilibrium relaxation, strictly true
only for $T_{substrate} = T_f$.  

It would seem natural to say that the enhanced mobility at the surface would
penetrate into the bulk at least on the length scales of the bulk dynamical
heterogeneity, as indeed it does.  This however results in a local relaxation
time which changes by many orders of magnitude over just a few particle
spacings.  This time scale gradient would be spread out via the dynamics of
mode coupling theory, diffusing the excess mobility deeper into the bulk.  In
first approximation, MCT has been shown to correlate dynamics on the length
scale $\xi_{MCT}$\cite{biroli.2004}, the predicted divergence of which at $T_c$
will be broken by the emergence of activated
events\cite{bhattacharyya.2005,bhattacharyya.2007}.  A simple equation
describing this gradient of relaxation rate incorporating activated events in
the framework of Bhattacharyya et al.\cite{bhattacharyya.2005} emerges

\begin{equation}
  \xi_{MCT}^2 \nabla^2 \tau^{-1}(z) - (\tau^{-1}(z) - \tau^{-1}_{local}(z) ) = 0.
\end{equation}

\noindent If we treat the mobility profile from purely activated dynamics,
$\tau^{-1}_{local}(z)$, as a set of boundary conditions at $z=0$ and $z=\infty$
(equivalent to coarse graining on the length scale $\xi_{CRR}$), then the
relaxation time, decaying smoothly from $\tau_{surf} \propto e^{-A_{surf} /
s_c}$ on the free surface to $\tau_{bulk}  \propto e^{-A_{bulk} / s_c} \ll
\tau_{surf}$ in the bulk on a length scale $\xi_{MCT}$, would follow

\begin{equation}
  \tau^{-1}(z) \approx \left( \tau^{-1}_{surf} - \tau^{-1}_{bulk} \right)
  e^{-z/\xi_{MCT}} + \tau^{-1}_{bulk}.
\end{equation}

\noindent The distance the excess mobility penetrates, $z^{*}$, is found by
comparing the magnitude of the two terms.  

\begin{equation}
  z^{*} = \xi_{MCT} \frac{A_{bulk} - A_{surf}}{s_c}
\end{equation}

\noindent We have shown that $A_{surf}$ can be as low as $\frac{1}{2}
A_{bulk}$, and that $A_{bulk}/s_c(T_g) \approx 40$ at the glass transition.
This gives a length scale $z^{*} \approx 20 \xi_{MCT}$ which can be much
larger than the bare dynamical correlation length.

At a depth $z<z^{*}$ from the surface the particle motion would freeze out when
$\tau^{-1}(z) \approx e^{-A_{surf} / s_c(T_g^{local})} e^{-z/\xi_{MCT}} \approx
\tau^{-1}_{bulk}(T_g)$, or when

\begin{equation}
  \frac{A_{surf}}{s_c(T_g^{local})} + \frac{z}{\xi_{MCT}} =
  \frac{A_{bulk}}{s_c(T_g)}
\end{equation}

\noindent We find that for the thinnest films local glass transition
temperature will be $T_g^{min} = T_g^{surf}$ as expected.  $T_g^{local}$ will
grow with distance from the free surface and be indistinguishable from the bulk
at a depth $z^{*}$. 

The surface properties of glasses are important in many technological and
biological contexts.  Enhanced surface mobility is relevant for adhesion,
friction, coatings, and nano-scale fabrication such as etching and lithography.
Supercooled water at the surface of proteins acts to enslave many protein
motions\cite{lubchenko.2005,frauenfelder.1991}.  Despite the influence of
static surface perturbations, we feel that the idealized treatment of the
surface mobility of glasses presented here can help in understanding these
phenomena.  The enhanced mobility at free surfaces will allow phase
transformations to occur at the surface that are kinetically impossible in
bulk.  De-vitrification often occurs at free glass surfaces, a fact of some
importance in geology\cite{marshall.1961} and archaeology\cite{richards.1936}.
This observation is naturally explained by the RFOT theory.

\bigskip 

\begin{acknowledgments} 
  Support from NSF grant CHE0317017 and NIH grant 5R01GM44557 is gratefully
  acknowledged.
\end{acknowledgments}

\bibliography{/people/jake/latex_files/jakes_biblio}
\end{document}